\documentclass[12pt]{article}
\usepackage{mathtext}
\usepackage{graphicx}
\usepackage[T2A]{fontenc}
\usepackage[utf8]{inputenc}
\usepackage{amsfonts}
\usepackage{amssymb}
\usepackage{amsmath}
\usepackage{pgfplots}
\usepackage{braket}
\usepackage{tikz}
\usepackage{float}
\begin{document}
\title{Qualitative model of a positive hydrogen peroxide ion in a thermal bath}

\author{Chen Ran$^{1}$, Yuri Ozhigov$^{1,2}$, You Jiangchuan$^1$\\
{\it 
    1.Moscow State University of M.V.Lomonosov,}\\{Faculty of Computational Mathematics and Cybernetics, Russia} 
    \\
    {\it 2. Valiev institute of physics and technology of Russian Academy of Sciences}
    \\
    }
\maketitle

\begin{abstract}
A qualitative model is proposed for a pair of atoms: oxygen and hydrogen in a single-mode optical cavity, bound by one valence electron and immersed in a thermal bath. The interaction of an electron with the cavity field depends on the state of the nuclei, which, in turn, is determined by the temperature of the phonon mode of the thermal bath. Computer simulation of the quantum dynamics of such a system shows the stable nature of the formation of both a stable molecular ion and a separate neutral oxygen atom and a positive hydrogen ion.
\end{abstract}

\section{Introduction}

The construction of a computer model of chemical reactions is an urgent task, the main obstacle to its solution is the problem of complexity. The association of atoms into a molecule implies the leakage of energy in the form of photons of various modes, and the exact consideration of such a problem even for fairly simple molecules seems to be very difficult due to the key role of the electromagnetic field in chemistry. Limiting the dimension of the problem by passing to a finite-dimensional model of quantum electrodynamics seems to be the best solution, especially since in applied problems the initial conditions for the approach of atoms can often be considered known and fixed (only processes such as explosions are an exception).

Most works in quantum chemistry are devoted to the exact calculation of binding energies (for example, \cite{M}) and conformations of molecules. Approaches to chemical dynamics classify the bonds between atoms without taking full account of the field dynamics (\cite{H}), for example, the study of various phase transitions in melts of elements by simplifying chemical bonds (\cite{H2}). In works of this type, the emphasis is on determining the binding energies and spectrum of molecules as accurately as possible, rather than on scaling the model to large ensembles; collective quantum effects (\cite{C}) are mainly studied in relation to already existing technologies. Non-Markovian dynamics is also studied in connection with the influence of phonons (see, for example, \cite{N}). The development of quantum electrodynamics in cavities (\cite{Phot}) makes it possible to implement more advanced operations in quantum electrodynamic systems, and also provides new ways to control ''field + matter'' systems. In polyatomic systems, in addition to the field, the thermal action of the medium in the form of thermal phonons with shorter wavelengths must also be included (see, for example, \cite{Phon}).

Meanwhile, computational models of chemistry at the quantum level have an independent value, since they help to build a new mathematical apparatus for the quantum physics of complex processes, which is currently not available. The obstacle here is the exponentially growing complexity of the real processes of the microcosm, which does not allow scaling standard models to hundreds and thousands of atoms even with the use of modern supercomputers. The pursuit of precision for simple diatomic molecules precludes obtaining even an approximate qualitative picture of the chemical scenario for many atoms.

The quantum description of a multimode electromagnetic field interacting with atoms requires the greatest computational resources.
Therefore, it seems appropriate to use the Jaynes-Cummings model (JC model - see \cite{JC} and \cite{Tav}), which reduces multimode dynamics to a simple single-mode picture of the interaction of a two-level atom or molecule in an optical cavity, with a single-mode field inside it in such a way that the frequency of the photon $\omega$ kept in the cavity coincides with a high accuracy with the frequency of excitation of the electron shell of the molecule. We will consider a diatomic molecule with one valence electron binding atoms, so that the ground state of the electron, denoted below by $|\Phi_0\rangle$, will bind atoms, and the excited state $|\Phi_1\rangle$ will loosen them.

Here, the interaction of the atom with the field inside the cavity is expressed in terms of the Hamiltonian
\begin{equation}
\begin{array}{ll}
&H_{JC}=H_0+H_{int},\ H_0=\hbar\omega a^+a+\hbar\omega \sigma^+\sigma,\ H_{int}=H_{int}^{RWA}+H_{int}^{strong},\\
&H_{int}^{RWA}=g(\sigma^+a+\sigma a^+),\ H_{int}^{strong}=g(\sigma^+a^++\sigma a).
\end{array}
\label{JC}
\end{equation}
where $a,a^+$ are the field operators for the annihilation and creation of a photon of the fundamental mode, and $\sigma,\sigma^+$ are the relaxation and excitation operators of the atom, respectively. For weak interactions characteristic of chemical reactions, we have $g/\omega\ll 10^{-2}$; in this case, one can omit the non-energy-conserving term $H_{int}^{strong}$ in the Hamiltonian, obtaining the energy operator in the form of a rotating wave approximation:
\begin{equation}
H_{JC}^{RWA}=\hbar\omega a^+a+\hbar\omega \sigma^+\sigma+g(\sigma^+a+\sigma a^+),
\label{RWA}
\end{equation}
which we will use in what follows, modifying it for the dynamics of atomic nuclei and electronic states in a molecule.

\section{Model description}

We will consider the process of association - dissociation of the hydrogen peroxide $OH^+$ molecule with one remote electron of the covalent bond, so that this bond is provided by only one remaining electron. We assume that this electron can tunnel between the nuclei of oxygen and hydrogen atoms in such a way that its two stationary states - the ground $|\Phi_0\rangle$ and the excited $|\Phi_1\rangle$ determine the binding strength of two atoms into a molecular ion.

According to the generally accepted notation of the molecular states of an electron (MO) as a linear combination of atomic orbitals (LCAO), we are looking for molecular states in the form
$$
|\Phi_0\rangle=\lambda^0_O|O\rangle+\lambda^0_H|H\rangle,\ |\Phi_1\rangle=\lambda^1_O|O\rangle+\lambda^1_H|H\rangle,
$$
where $|O\rangle,\ |H\rangle$ are the positions of the electron at the oxygen and hydrogen atoms, respectively, and the amplitudes $\lambda$ have the property that
$$
|\lambda^0_O|>|\lambda^0_H|,\ |\lambda^1_O|<|\lambda^1_H|.
$$
This follows from the method of finding molecular states as eigenvectors of the tunneling operator $H_{tun}=\frac{a}{2}(I-\sigma_z)-g\sigma_x$, where $a>0$ is the potential difference of an electron at oxygen and hydrogen, and $g>0$ is inversely proportional to the height of the potential barrier between these atoms with respect to the covalent electron. A simple calculation gives the value $\delta E=\sqrt{a^2+4g^2}$ for the energy gap between molecular orbitals. With the removal of atomic nuclei, the conditional value of the barrier $1/g$ increases and the gap decreases; in this case, the states $|\Phi_0\rangle,\ |\Phi_1\rangle$ also change - we will neglect the last effect, especially since we will not take into account these orbitals in the basic states of the model. The distance between atomic nuclei will be taken into account only in the form of influence on the energies of interaction with the field of these states. In an optical cavity in which atoms are located, the interaction energy is expressed as
$$
g=\sqrt{\frac{\hbar\omega}{V}}E(x)d_{0,1}
$$
where $d_{0,1}$ is the matrix element of the dipole transition between orbitals, which changes more slowly than the frequency $\omega$, so that for a close distance between the nuclei, the coefficient $g$ will be higher than for a far one. Variation in the $\omega$ frequency will result in a fast leakage of the transition photon between $|\Phi_0\rangle$ and $|\Phi_1\rangle$, so that our system being in a cavity is not a defining detail of the model.

Our goal is: a) to establish the nature of electron oscillations between the positions $|O\rangle$ and $|H\rangle$ and its qualitative dependence on the positions of atomic nuclei, and b) the dependence of the probabilities of the reaction channels $O+H^+\leftrightarrow OH ^+$ on the initial conditions and model parameters.

Note that the oscillations for point a) are determined not by the $H_{tun}$ operator, but by the full Hamiltonian $H_{JC}^{RWA}$, that is, by the interaction of molecular orbitals with the electromagnetic field.

We assume that the nuclei of $O$ and $H$ atoms have very low kinetic energy, so that a covalent bond between atoms can exist, although it is very weak. The dynamics of nuclei is influenced not only by the state of the electron that binds them, but also by the phonons of the thermal bath into which the pair of atoms under consideration is immersed. The interaction of nuclei with phonons can be expressed by the Hamiltonian
$$
H_{term}=g_c(b^+\sigma_c+b\sigma_c^+)
$$
where $b,b^+$ are the phonon annihilation and creation operators, and $\sigma_c,\sigma_c^+$ are the relaxation and excitation operators of nuclear dynamics.
However, the inclusion of the number of phonons directly in the basis states would lead to an unjustified increase in the required memory, so we restrict ourselves to the assumption that the average number of phonons $m_{ev}$ changes very slowly compared to the oscillations of nuclei, so that we can write the interaction $H_{term} $ as $H_{term}=g_c(\sigma_c^++\sigma)$, where $g_c=g_0\sqrt{m_{ev}}$.

To find the dependence of the coefficient $g_c$ on the temperature of the phonon bath, we turn to the result of \cite{Oz}, where it was proved that thermal stabilization in the Jaynes-Cummings model in the RWA approximation for the exchange of bosons between the medium and the cavity gives a state of the bosonic field of the form
$$
G_T=c\cdot\sum\limits_{n=0}^\infty e^{-\frac{\hbar\omega_c n}{KT}}|n\rangle\langle n|
$$
where $\omega_c$ is the frequency of the exchange bosons between the medium and the cavity, $T$ is the temperature inside the cavity, $K$ is the Boltzmann constant, and the coefficient $\mu=e^{-\frac{\hbar\omega_c}{KT} }$ coincides with the partial $\mu=\gamma_{in}/\gamma_{out}$ - the ratio of the amplitude of the inflow and outflow of bosons into the cavity. Knowing these amplitudes, one can find the temperature inside the cavity.

So, the basic states for atomic dynamics will have the form
$$
|n\rangle|el\rangle|nuc\rangle
$$
where $n=0.1$ is the number of free photons in the cavity, $el=0.1$ is the state of the valence electron - ground and excited respectively, $nuc$ is the ground and excited dynamic state of atomic nuclei. For nuclei, we will conditionally assume that $nuc=0$ means the ground state of the nuclei when they are close and the presence of a covalent bond is determined by the state of the electron ($|\Phi_0\rangle$ is a bond, $|\Phi_1\rangle$ - no connection), and $nuc=1$ means that the nuclei are far away, when their kinetic energy is greater and there is no covalent bond. The coefficient $g_c$ of the interaction between the dynamics of nuclei and phonons will depend both on the temperature $T$ of the phonon mode inside the cavity and on the state of the electron: for the ground state of the electron it will be smaller.

Finally, we will assume that the photon of the electronic excitation is quickly removed from the cavity, that is, the ''temperature'' of the photons tends to zero. Under these assumptions, to find the finite probabilities of the association-dissociation reaction channels, we will have a Hamiltonian of the form
\begin{equation}
\label{ham}
H=H_0+g_c(el)(\sigma_c^++\sigma_c)+g_{el}(nuc)(a^+\sigma_{el}+a\sigma_{el}^+),\ H_0=\hbar\omega_c\sigma_c^+\sigma_c+\hbar\omega_{el}\sigma^+_{el}\sigma_{el}+\hbar\omega_{el}a^+a.
\end{equation}

\section{Results of computer simulation}

   So, we have only these 6 basic states:
    \begin{equation*}
        |000\rangle,\ |001\rangle,\ |010\rangle,\ |011\rangle,\ |100\rangle,\ |101\rangle. 
    \end{equation*}
    
    We write the Hamiltonian \eqref{ham} in the following matrix form:
	\begin{equation*}	
		\begin{array}{l|ccccccc} 
			& |000\rangle & |001\rangle  & |010\rangle & |011\rangle & |100\rangle  & |101\rangle\\ 
			\hline 
			|000\rangle & 0    & g_{c}(0)      & 0     & 0   & 0     & 0 \\ 
			|001\rangle & g_{c}(0)    & \hbar\omega_{с}      & 0     & 0  & 0     & 0 \\ 
			|010\rangle & 0    & 0      & \hbar\omega_{el}   & g_{c}(1) & g_{el}(0) & 0 \\ 
			|011\rangle & 0    & 0      & g_{c}(1)     & \hbar\omega_{el} + \hbar\omega_{с} & 0 & g_{el}(1) \\ 
			|100\rangle & 0    & 0      & g_{el}(0)     & 0  &\hbar\omega_{el}  & g_{c}(0)\\
			|101\rangle & 0    & 0      & 0     & g_{el}(1) & g_{c}(0) & \hbar\omega_{el} + \hbar\omega_{с}
		\end{array}	
	\end{equation*}

    \begin{equation*}
		g_{el}(0) \gg g_{el}(1) > g_c(1) \gg g_c(0)
	\end{equation*}
 
   The relaxation of the atomic system will be represented as a solution to the quantum master equation (QE):
   
	\begin{equation*}
		i\hbar \dot \rho = [H, \rho] + i L(\rho)
	\end{equation*}
	
	where $L(\rho)$ is the Lindblad operator, which in the general case has the form:
    
	\begin{equation*}
		L(\rho) = \sum_{j=1}^N \gamma_j \left( A_j \rho A_j^+ - \frac{1}{2}\left(\rho A_j^+A_j + A_j^+A_j \rho \right)\right)
	\end{equation*}

    Since the Lindblad operator gives a small addition compared to the unitary dynamics, using the Euler method to solve it is quite sufficient. In our case, it is convenient to represent it as an iteration of two steps:
    
    \begin{equation*}
        \begin{aligned}
            & 1. \ \rho'(t + \Delta t) = e^{-i H \Delta t} \rho e^{i H \Delta t}\\
            & 2. \ \rho(t + \Delta t) = \rho'(t + \Delta t) + L(\rho'(t + \Delta t)) \Delta t
        \end{aligned}
    \end{equation*}

    To simplify the calculation, we used the following values of the parameters that satisfy the specified conditions:
	\begin{equation*}
		\begin{array}{ccc}
		     &\hbar = 1,  \Delta t = 0.01, \mu = \frac{\gamma_{in}}{\gamma_{out}} = 0.4,\\
            & g_{el}(0) = 6 * 10 ^ 7  \gg g_{el}(1) = 6 * 10 ^ 4 > g_{c}(1) \gg g_{c}(0), \\
             & g_{c}(1) = 4 \cdot 10 ^ 3 \cdot e^{-0.1t}\to 0, g_{c}(0) = 4  \cdot e^{-0.1t} \to 0,
		\end{array}
		\label{fig:parameter}
	\end{equation*}
    where $t = iteration*\Delta t, iteration \in [0,6000]$ $iteration$ is the number of iterations.

The calculation results give the following probability distributions of the basic states of the system depending on time:

    \begin{figure}[H]
		\centering
		\begin{minipage}[t]{0.49\linewidth}	\includegraphics[width=3in]{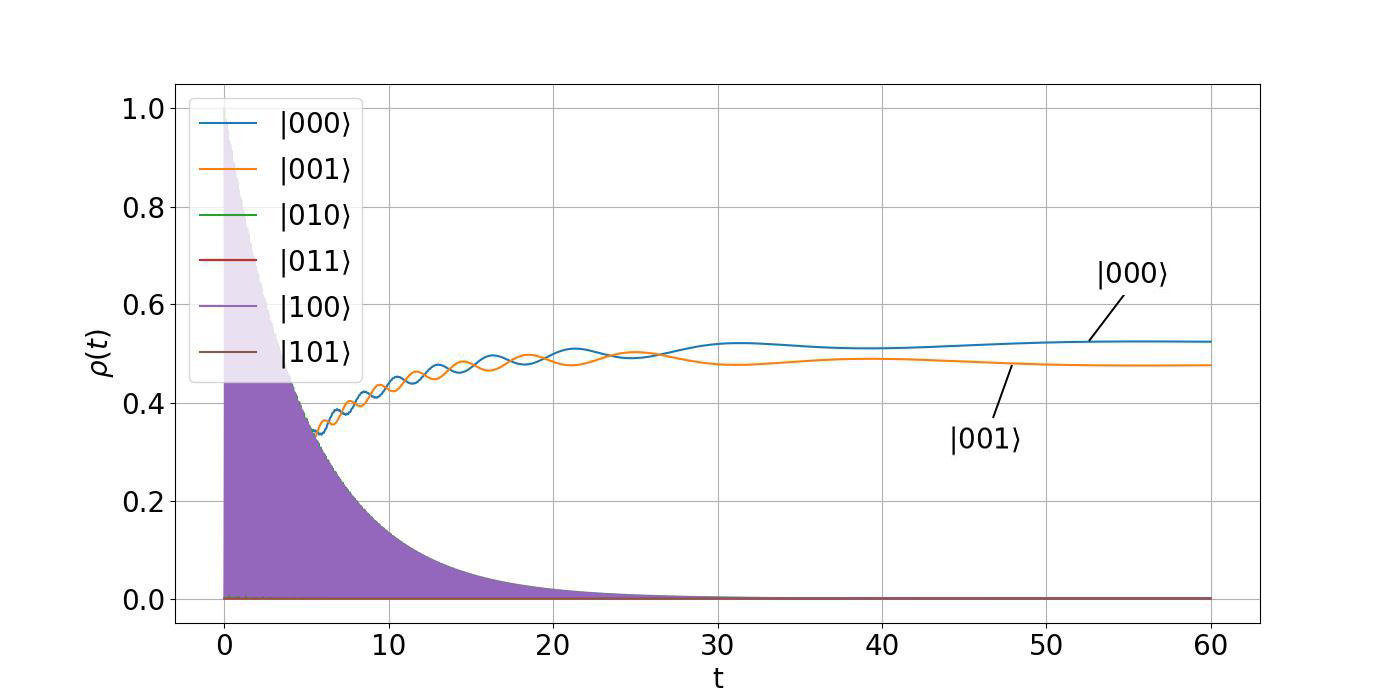}
			\caption{Initial state: $|010\rangle$}\label{fig:image_OH_010_04.png} 
		\end{minipage}
        \begin{minipage}[t]{0.49\linewidth}
			\centering\includegraphics[width=3in]{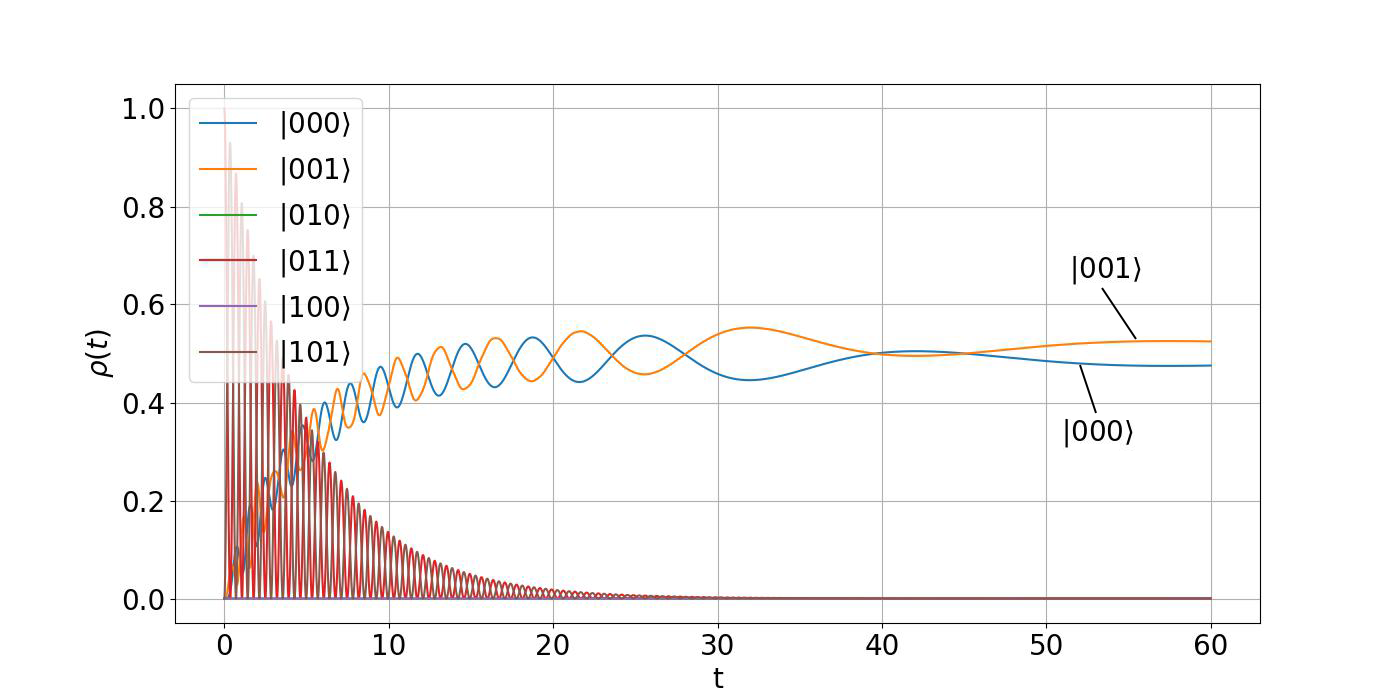}
			\caption{Initial state: $|011\rangle$}\label{fig:image_OH_011_04.png} 
		\end{minipage}
        $\omega_{el} =  2\omega_{c} = 0.4$
	\end{figure} 

    \begin{figure}[H]
		\centering
		\begin{minipage}[t]{0.49\linewidth}	\includegraphics[width=3in]{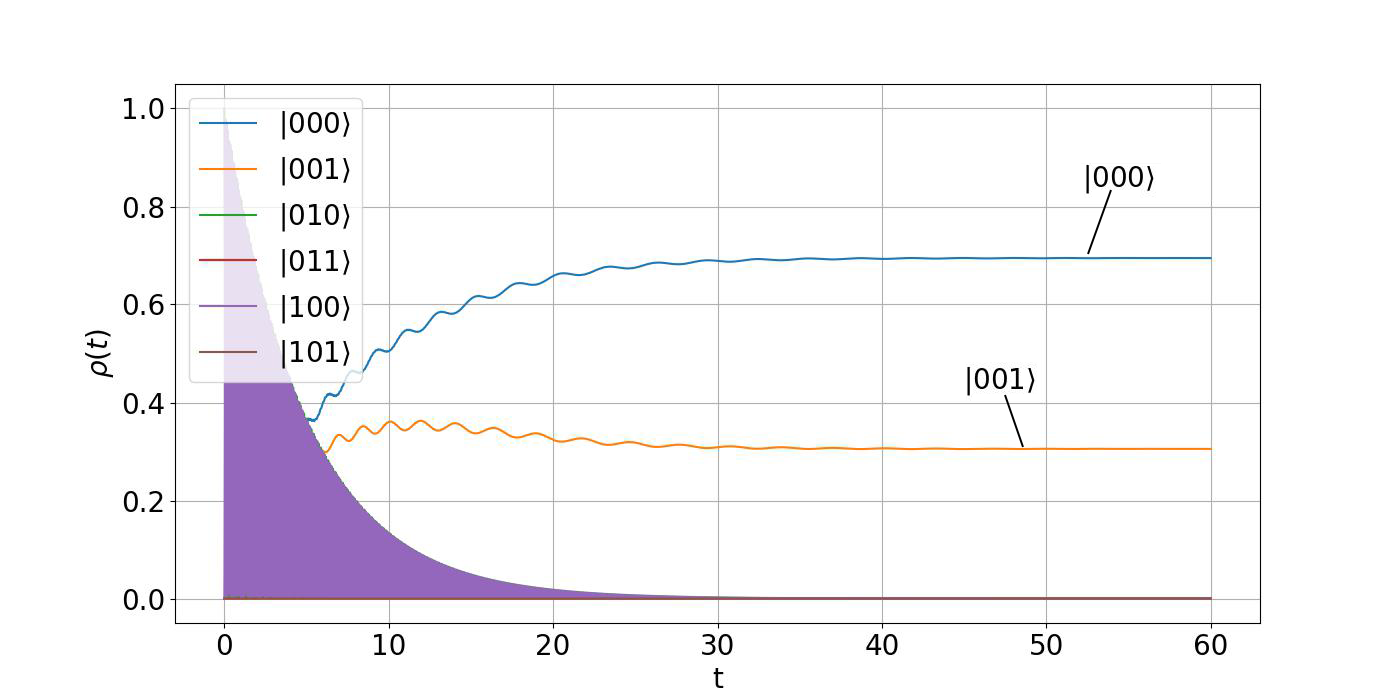}
			\caption{Initial state: $|010\rangle$}\label{fig:image_OH_010_4.png} 
		\end{minipage}
        \begin{minipage}[t]{0.49\linewidth}
			\centering\includegraphics[width=3in]{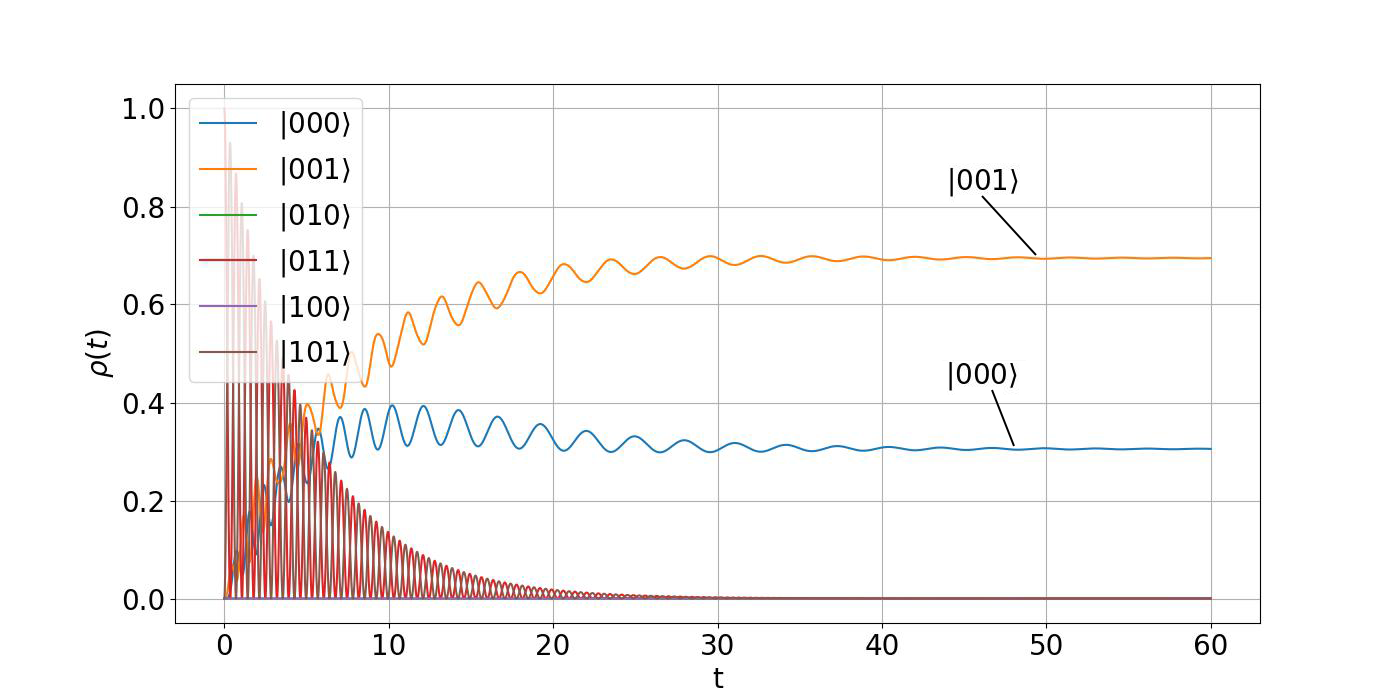}
			\caption{Initial state: $|011\rangle$}\label{fig:image_OH_011_4.png} 
		\end{minipage}
        $\omega_{el} =  2\omega_{c} = 4$
	\end{figure} 
 
    \begin{figure}[H]
		\centering
		\begin{minipage}[t]{0.49\linewidth}	\includegraphics[width=3in]{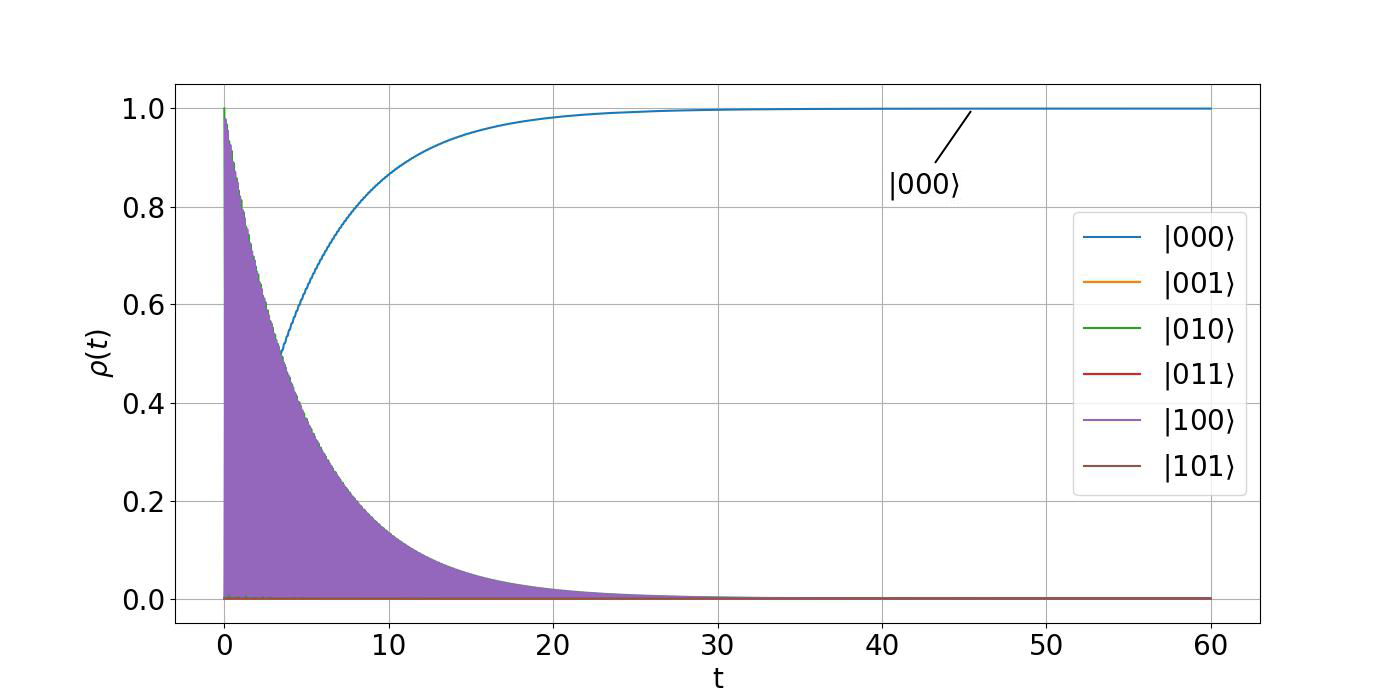}
			\caption{Initial state: $|010\rangle$}\label{fig:image_OH_010_400.png} 
		\end{minipage}
        \begin{minipage}[t]{0.49\linewidth}
			\centering\includegraphics[width=3in]{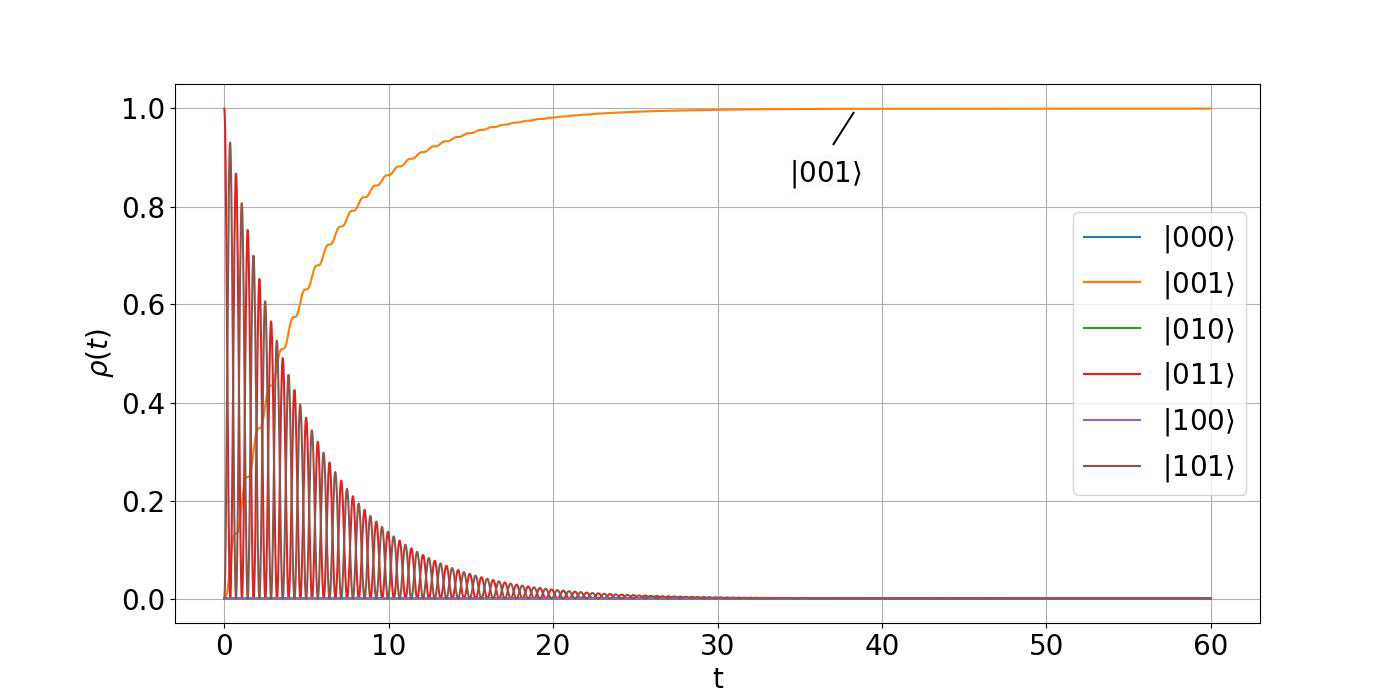}
			\caption{Initial state: $|011\rangle$}\label{fig:image_OH_011_400.png} 
		\end{minipage}
        $\omega_{el} =  2\omega_{c} = 400$
	\end{figure} 

    The resulting probability distributions thus strongly depend not only on the initial conditions, but also on the absolute values of the frequencies of electronic transitions and thermal phonons (figures \ref{fig:image_OH_010_04.png} and \ref{fig:image_OH_010_4.png}). In particular, for low frequencies, when approaching the applicability limit of the rotating wave approximation, noticeable oscillations in the probability of forming a covalent bond are observed (figures \ref{fig:image_OH_011_04.png} and \ref{fig:image_OH_011_4.png}). For high frequencies of electronic transitions, where the rotating wave approximation is very accurate, a smooth behavior of the association probability into a molecule is observed (Fig. \ref{fig:image_OH_010_400.png} and \ref{fig:image_OH_011_400.png}).

\section{Transition frequencies of electrons to potential wells of different depths.}

Let us now consider in more detail the dynamics of an electron in a covalent bond formed between conventional oxygen and hydrogen atoms. We neglect the effects associated with the shape of the electron clouds of valence atoms, representing the valence electron as a particle tunneling between two potential wells, as shown in the figure \ref{psi0-1OH.png}:

	\begin{figure}[H]
		\centering
		\includegraphics[height=2in,width=2in]{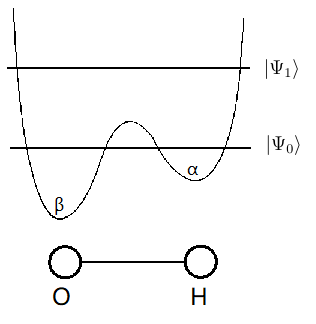}
		\caption{Basic and excited orbitals}
		\label{psi0-1OH.png} 
	\end{figure} 

Here, the depths of potential wells are related by the condition: $\alpha>\beta$, since the oxygen atom attracts the electron much more strongly than the hydrogen atom. The eigenstates of the electron will have the form

$$
 |\Psi_0\rangle = \frac{\alpha|O\rangle+\beta|H\rangle}{\sqrt{|\alpha|^2+|\beta|^2}},\    
 |\Psi_1\rangle = \frac{-\beta|O\rangle+\alpha|H\rangle}{\sqrt{|\alpha|^2+|\beta|^2}},              
$$

where $|O\rangle$ and $|H\rangle$ are the orbitals of the excited state of oxygen and hydrogen atoms, respectively. $|\Psi_0\rangle$ and $|\Psi_1\rangle$ are hybrid (molecular) orbitals corresponding to the ground and excited states, respectively.

The Hamiltonian of our problem has the form:
$H=\hbar\omega a_{\omega}^+a_{\omega}+\hbar\omega\sigma_{\omega}^+\sigma_{\omega}+g_{mol}(a_{\omega}^+\sigma_{\omega}+a_{\omega}\sigma_{\omega}^+),$

where relaxation operators - excitations, as well as field operators, refer already to the $OH$ molecule. Thus, $a_{\omega}\sigma_{\omega}^+$ reflects the absorption of a photon by an electron during the transition from the ground state to an excited one, while $a_{\omega}^+\sigma_{\omega}$ reflects the opposite, emission of a photon during the transition from an excited state to the ground state. $g_{mol}=\frac{1}{L}$ - is the energy of electron interaction with the field, $L$ is the conditional distance between atoms.
The matrix representation of the Hamiltonian has the form
$$
H = 
\begin{array}{l|cccc} 
      &|0\rangle|\Psi_0\rangle  &|0\rangle|\Psi_1\rangle &|1\rangle|\Psi_0\rangle  &|1\rangle|\Psi_1\rangle\\ 
    \hline 
|0\rangle|\Psi_0\rangle & 0 & 0    & 0    & 0\\
|0\rangle|\Psi_1\rangle & 0 & \hbar\omega    & g_{mol} & 0\\
|1\rangle|\Psi_0\rangle & 0 & g_{mol} & \hbar\omega    & 0\\
|1\rangle|\Psi_1\rangle & 0 & 0    & 0    & 2\hbar\omega\\
\end{array} 
$$
Let us find analytically the solution of the Schrödinger equation using the standard formula

$|\Psi(t)\rangle=\sum_j e^{\frac{-i}{\hbar}E_jt}\lambda_j|\psi_j\rangle,$ 
where $\lambda_j=\langle\Psi(0)|\psi_j\rangle$,$|\psi_j\rangle$ -
eigenvector of the Hamiltonian H, $E_j$ is the corresponding eigenvalue, and $|\Psi(0)\rangle=|0\rangle_{ph}|O\rangle=\frac{\alpha|0\rangle_{ph}|\ Psi_0\rangle-\beta|0\rangle_{ph}|\Psi_1\rangle}{\sqrt{\alpha^2+\beta^2}}$. We have:
$$
E_1=\hbar\omega-g_{mol},|\psi_1\rangle=
\left(
\begin{array}{l} 
0  \\
\frac{1}{\sqrt{2}}  \\
\frac{-1}{\sqrt{2}}  \\
0  
\end{array}
\right),
E_2=\hbar\omega+g_{mol},|\psi_2\rangle=
\left(
\begin{array}{l} 
0  \\
\frac{1}{\sqrt{2}}  \\
\frac{1}{\sqrt{2}}  \\
0  
\end{array}
\right),
$$
$$
E_3=2\hbar\omega,|\psi_3\rangle=
\left(
\begin{array}{l} 
0  \\
0  \\
0  \\
1  
\end{array}
\right) 
$$

Where

$|\Psi(t)\rangle=\frac{2\alpha^2+\beta^2(e^{iE_1t}+e^{iE_2t})}{2(\alpha^2+\beta^2)}|0\rangle|O\rangle+\frac{\alpha\beta(e^{iE_1t}-e^{iE_2t})}{2(\alpha^2+\beta^2)}|1\rangle|O\rangle+\frac{2\alpha\beta-\alpha\beta(e^{iE_1t}+e^{iE_2t})}{2(\alpha^2+\beta^2)}|0\rangle|H\rangle+\frac{-\beta^2(e^{iE_1t}-e^{iE_2t})}{2(\alpha^2+\beta^2)}|1\rangle|H\rangle$

~\\
After simplification, we obtain the probability $P(|O\rangle)$ of the appearance of an electron in the first well (on the first atom):

$P(|O\rangle)=P(|0\rangle|O\rangle)+P(|1\rangle|O\rangle)=\frac{1}{4(\alpha^2+\beta^2)^2}(|2\alpha^2+\beta^2(e^{iE_1t}+e^{iE_2t})|^2
+\alpha^2\beta^2|e^{iE_1t}-e^{iE_2t}|^2)=\frac{1}{4(\alpha^2+\beta^2)^2}(C+\beta^2\frac{\beta^2-\alpha^2}{2}\cos{2g_{mol} t}+2\alpha^2\beta^2\cos{\hbar\omega t}\cos{g_{mol} t})$,

here $C=\alpha^4+\frac{\beta^4}{2}+\frac{\alpha^2\beta^2}{2}$. The following graphs represent the resulting solution.
	\begin{figure}[H]
		\centering
		\includegraphics[width=6.2in]{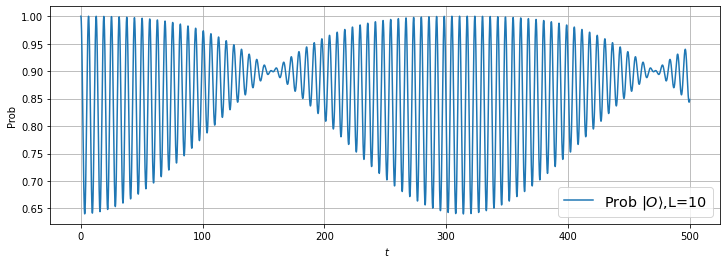}
		\caption{The probability of an electron appearing near an oxygen atom when there is no leakage of photons in the resonator.}
		\label{fig:OH1e-1.png} 
	\end{figure} 

	\begin{figure}[H]
		\centering
		\includegraphics[width=6.2in]{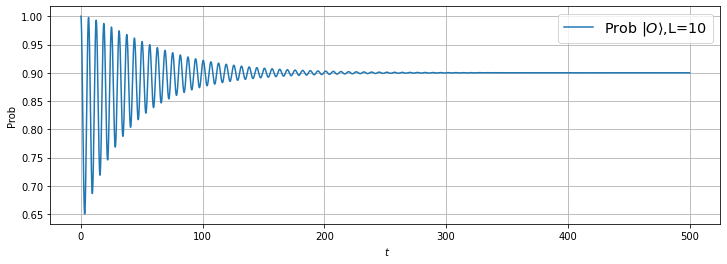}
		\caption{The probability of an electron appearing near an oxygen atom is a photon leakage (decoherence $A=a$ in COA)}
		\label{fig:OH1e-2.png} 
	\end{figure} 

where $P(|O\rangle)$ tends to $y=\alpha^2$. We see that electrons during relaxation have a high probability of being in deeper potential wells, which was to be expected.

\section{Conclusions}

     We have proposed a simple model for studying the processes of association and dissociation of two different atoms - oxygen and hydrogen, the covalent bond between which is carried out by one electron. Our model includes the interaction of two molecular orbitals of an electron with the field, the simplest version of the motion of nuclei, and the influence of the temperature factor of the medium on this motion.

     It has been established that, under different initial conditions, the predominant reaction channel will be the formation of the $O,H^+$ pair. In this case, the specific probability of reaction channels with the formation of such a pair and the pair $O^+,H$ depends on the initial condition. An interesting dependence of the probabilities of channels with these reaction products on the rate of cooling of the environment has also been established.

     These results show that our simple model can be extended to ensembles of hundreds of atoms, taking into account both the electromagnetic field and environmental thermal phonons. Such a scaling can provide a good tool for searching for important multiparticle effects of a quantum nature in chemistry.

\end{document}